\documentclass[superscriptaddress,aps,prl,twocolumn,showpacs,preprintnumbers,floatfix]{revtex4-1}

\usepackage[cmex10]{amsmath}
\usepackage{amssymb}
\usepackage{graphicx}
\usepackage{dcolumn}
\usepackage{bm}

\newcommand{\ket}[1]{|#1\rangle}

\begin{document}

\title{Detection of ultra-high resonance contrast in vapor cell atomic clocks}

\author{Jinda Lin}
\affiliation{Key Laboratory for Quantum Optics, Center for Cold Atom Physics, Shanghai Institute
of Optics and Fine Mechanics, Chinese Academy of Sciences, Shanghai 201800, China}

\author{Jianliao Deng}
\affiliation{Key Laboratory for Quantum Optics, Center for Cold Atom Physics, Shanghai Institute
of Optics and Fine Mechanics, Chinese Academy of Sciences, Shanghai 201800, China}

\author{Yisheng Ma}
\affiliation{Key Laboratory for Quantum Optics, Center for Cold Atom Physics, Shanghai Institute
of Optics and Fine Mechanics, Chinese Academy of Sciences, Shanghai 201800, China}

\author{Huijuan He}
\affiliation{Key Laboratory for Quantum Optics, Center for Cold Atom Physics, Shanghai Institute
of Optics and Fine Mechanics, Chinese Academy of Sciences, Shanghai 201800, China}

\author{Yuzhu Wang}
\email{yzwang@mail.shcnc.ac.cn}
\affiliation{Key Laboratory for Quantum Optics, Center for Cold Atom Physics, Shanghai Institute
of Optics and Fine Mechanics, Chinese Academy of Sciences, Shanghai 201800, China}

\date{\today}

\begin{abstract}
 We propose and demonstrate a novel detection scheme of clock signals and obtain an ultra-high resonance contrast above 90\%. The precision of the signal's detection and the signal-to-noise ratio (SNR) of atomic clock signal is improved remarkably. The frequency stability in terms of Allan deviation has been improved by an order for the new detection under the equivalent conditions. We also investigate density effect which produces the splitting of the transmission peak and consequently a narrower linewidth of Ramsey fringes.
\end{abstract}


\maketitle

Atomic clocks play a key role in many aspects of modern life, especially in time-keeping, synchronization, and communication\cite{Vanier1989, Vanier2007}. Although atomic clocks operated in optical frequency regime have attracted increasing interest as regards the next frequency standards, microwave atomic clocks are still irreplaceable in the near future because of its mature technical implementation. Active efforts have been contributed to improve clock performance, especially the contrast and signal-to-noise ratio (SNR), which are defined in the Allan deviation \cite{Allan1966}. In state of the art, the frequency stability of atomic clock is limited by shot noise \cite{Zibrov2010}. Over years, several techniques, such as coherent population trapping (CPT) resonance \cite{Svenja2004,Vanier2003}, N-resonance \cite{Zibrov2005,Novikova2006} and electromagnetically induced transparency (EIT) frequency standards \cite{2Vanier2003}, have been proposed to eliminate systemic factors that deteriorate the clock performance and improve short-term frequency stability of the clocks. To further cancel the first-order light shift, pulsed optically pumped (POP) \cite{Godone2004} and pulse coherence storage (PCS) \cite{Yan2009} clocks have been also investigated, in which microwave and optical fields interact with the atoms successively in timing sequence \cite{Ardit1964}. In POP and CPT clocks, one can obtain the information of the clock by measuring the absorption of transmitted probe lasers. To date, the contrast up to 30\% have been reported \cite{Zibrov2010,Micalizio2012}. The achievement of frequency stability of few units of 10$^{-15}$ has been reported in a POP vapor cell clock \cite{Micalizio2012}.

In this paper, we develop a novel detection technique to improve clock performance and report an ultra-high contrast of above 90\%. In previous studies, it has been shown that the dispersive detection has the advantages of high sensibility and high SNR over the absorption detection \cite{Budker2002,Zigdon2010,Jackson2009}. However, to our knowledge, it is the first time that the dispersive detection has been realized in microwave clocks and an amazing ultra-high contrast is reached.

\begin{figure}
\includegraphics[width=8cm]{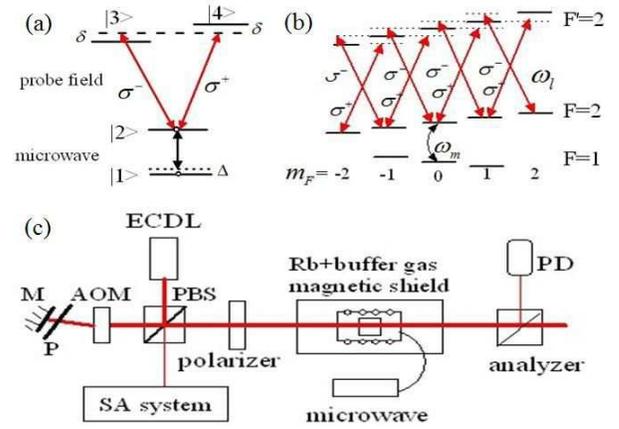}\\
\caption{(Color online)Dispersive detection of the clock transition. (a) Four-level model of dispersive detection. (b) Energy levels for $^{87}$Rb $D_1$ line. (c) Typical experimental setup for the detection of the reference transition signal. $\omega_l$: light circular frequency, $\omega_m$ :microwave circular frequency. P: quarter wave plate, M: reflector, PBS: polarized beam splitter, PD : photodiode.\label{levels}}
\end{figure}

\textit{Dispersive detection of the clock transition}---
Our scheme of the dispersive detection can be illustrated with a four-level atomic model as shown in Fig.\ref{levels}(a). The microwave field drives the clock transition from $\ket{1}$ to $\ket{2}$. A linear-polarized probe light transmits an atomic sample, and the forward scattering light is detected after the polarization of the light field is rotated \cite{Budker2002,Gawlik1974,Horrom2011,Siddons2009,Lvovsky2009}. In the presence of a weak longitudinal magnetic field, the left- and right-circular components of the  probe light couple the detection transition from $\ket{2}$ to the states $\ket{3}$ and $\ket{4}$, respectively. Specially, when probe laser is exactly resonant with the detection transition, the detunings of two components with different circular polarizations are equal with opposite signs. The transmission of resonant optical field through the atomic vapor positioned between two crossed polarizers is given by \cite{Gawlik1979}:
\begin{equation}\label{eq:transmission}
    I_s = |E_0|^2 \exp^{-[(\alpha_{+} + \alpha_{-}) k L]} \sin^2 \left[ \frac{(n_{+} - n_{-}) k L}{2} \right],
\end{equation}
where $k$ is the incident wave vector, $E_0$ is the incident electric field, $n_{\pm} \propto \text{Re} (\rho_{\pm})$ and $\alpha_{\pm} \propto \text{Im} (\rho_{\pm})$ are the refractive index and absorption factors for  $\sigma^{\pm}$.  $\rho_{\pm}$ for $\sigma^{\pm}$ can be calculated from the optical Block equation of the density matrix as follows \cite{Arimondo1996}:
\begin{equation}\label{eq:master}
    \frac{d \rho}{dt} = \frac{1}{i \hbar} [H, \rho] + {\Gamma \rho},
\end{equation}
where $H$ is the Hamiltonian of the four-level model, and $\Gamma$ is the relaxation matrix. After a bit of algebra, we obtain the steady-state solution of the density matrix elements. The calculation of $\rho_{\pm}$ turns out:
\begin{equation}\label{eq:rho}
    \rho_{\pm} \approx  \frac{-i\Omega_l /2}{\frac{\Gamma^*}{2} \pm i \delta_l} \cdot \frac{\Omega_l^2/2}{2\gamma_l \Gamma^* + \Omega^2_l} \cdot \frac{\Omega_m^2}{(\gamma_1 + \frac{\Omega_l^2}{\Gamma^*})(\gamma_2 + \frac{\Omega_l^2}{\Gamma^*}) + \Delta^2 + \Omega_m^2},
\end{equation}
where $\Omega_{l,m}$ are the Rabi frequency of laser and microwave. $\delta_l, \Delta$ are the detunings of laser and microwave. $\Gamma^* = 3\times 10^9$ s$^{-1}$, $\gamma_{1,2}=300s^{-1}$ are the relaxation rate of excite states, ground state population, and ground state coherence relaxation, respectively. \\

The exponential part in equation (2) corresponds with the absorption and the last term describes the birefringence. The birefringence and the absorption scales differently with the atomic population of $\ket{2}$, which is dependent on the microwave excitation on the clock transition. The transmitted intensity of the forward scattering light $I_s$ reflects the reference transition signal of the clock transition. While the microwave excitation on the clock transition is weak, which means fewer atoms are excited to $\ket{2}$, the optical absorption of the medium is negligible but the rotation is still detectable. In this case, $I_s$ is linearly dependent on the atomic population of $\ket{2}$. As the microwave excitation becomes strong, more and more atoms are populated in $|2>$, and the optical absorption of the medium becomes dominant. Now the dependence of $I_s$ on the atomic population excited by the microwave field is nonlinear rather than linear. In the different regimes of atomic density of the sample, the dispersion detection has different characteristic, which will be described bellow.

\textit{Measurement of the resonance contrast of the clock signal}---
The detection is implemented in $D_1$ line of $^{87}$Rb with laser frequency tuned to the $S_{1/2} - P_{1/2}$ transition. The specific levels involved in the experiment are shown in Fig.\ref{levels}(b). A typical experimental setup is shown in Fig. \ref{levels}(c). The incident optical field is derived from an external cavity diode laser (ECDL). The laser frequency is stabilized and tuned to be resonant with the 795nm $\ket{F=2} \rightarrow \ket{F'=2}$ transition. The diameter of laser beam is about 5mm and the maximum intensity is $7.1mW/cm^2$. An acoustic-optical modulator (AOM) is used to shift the light frequency by 160MHz to cancel the buffer gas induced level shift and to act as an optical switch to generate light pulses. The light field propagates along the direction of the constant longitudinal magnetic field, and becomes linearly polarized light after passing through a linear polarizer. An analyzer adjusted orthogonal to the polarizer ($\phi=90^\circ$) blocks the optical field background away from the photodiode (PD). When the microwave field is far off-resonant with the transition between $\ket{F=1,m_F=0}$ and $\ket{F=2,m_F=0}$, the reading of PD is mainly determined by the extinction ratio of the total system (polarizer, analyzer and vapor cell).

We perform the experiments in a quartz made cylindrical cell which is 20mm in length and 30mm in diameter. The cell contains isotopically enriched $^{87}$Rb vapor and mixture buffer gases of Ar (15.5Torr) and N2 (9.5Torr). The buffer gas is added to reduce atomic decoherence induced by collision with the walls of the cell and to suppress Doppler broadening by Dicke effect \cite{Vanier2007,Gondone2000}. The whole cell is heated to about 65$^\circ$. The cell is placed inside a microwave cavity whose Q factor is about 1000. The cavity operates on the TE$_{011}$ mode and is tuned to 6.834 GHz. A solenoid around microwave cavity produces a constant longitudinal magnetic field. The magnetic field of 20mG is used to lift the Zeeman sublevels apart. A microwave field operates on the clock transition between $\ket{F=1,m_F=0}$ and $\ket{F=2,m_F=0}$. The cell, the cavity and the solenoid are placed inside a three layers of cylindrical $\mu$-metal shield to isolate them from external magnetic fields.

\begin{figure}
\includegraphics[width=8cm]{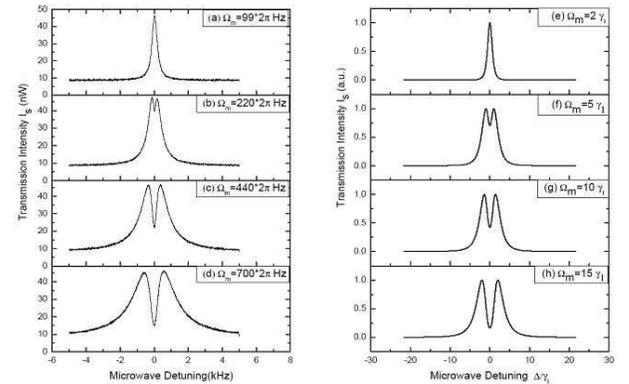}\\
\caption{Experimental detection (a-d) and theoretical calculations (e-h) of the transmitted light intensity  $I_s$.  Left panel is experimental results: the probe intensity was 45 $\mu$W/cm$^2$ and microwave Rabi frequency  $\Omega_m$ is $99\times 2\pi$, $220\times 2\pi$ , $440\times 2\pi$ , and $700\times 2\pi$  Hz, respectively. Right panel is the simulation results:  $\Omega_l = 10^4 \gamma$, and microwave Rabi frequency  $\Omega_m$ is $2\gamma_1$; $5\gamma_1$; $10\gamma_1$; and $15\gamma_1$, respectively. The peak of the theoretical plots is matched to the experimental maximum.\label{split}}
\end{figure}

The microwave excitation of the clock transition is directly determined by the microwave detuning, and consequently the transmission of the probe light depends remarkably on the microwave detuning and exhibits different profiles in the regimes of the low and high atomic density. With a low atomic density and a weak probe field, the transmitted light intensity increases as the microwave detuning decreases and reaches its maximum at the exact resonance. In this case, the forward scattering signal exhibits the linear magneto-optical effect, in other words, the detected signal exactly describes microwave interrogation of the two-level clock transition. The transmission's dependence on the microwave detuning is shown in Fig.\ref{split}(a). Here we observed an ultra-high resonance contrast $C=\frac{I_{s,max}-I_{s,min}}{I_{s,max}} \approx 80\%$. $I_{s,min}$ is the residual light background as a result of finite extinction ratio of the analyzer.

\begin{figure}
\includegraphics[width=8cm]{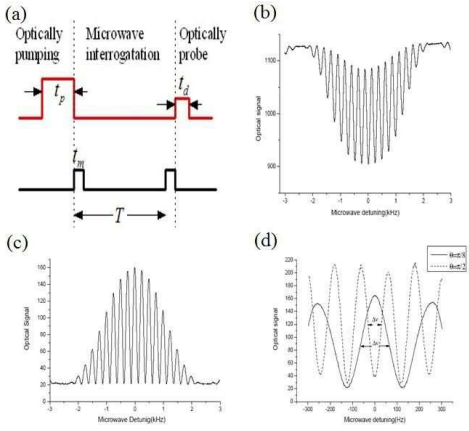}\\
\caption{(Color online) POP performance with the dispersive detection. (a) Experimental timing sequence: $t_p=2ms , t_m=0.4ms, t_d=0.2ms, and T=4ms$  are the optical pumping time, microwave pulse time, detection time, and Ramsey time respectively. (b) Dispersive detection of Ramsey fringe. (c) Absorption detection of Ramsey fringe. (d) Central zone of Ramsey fringes with pulse areas $\theta=\pi/8$ ,$\Delta\nu=128$  Hz (solid line) and  $\theta=\pi/2$ ,$\Delta\nu=64$ Hz(dashed line).\label{Ramsey}}
\end{figure}

\begin{figure}
\includegraphics[width=8cm]{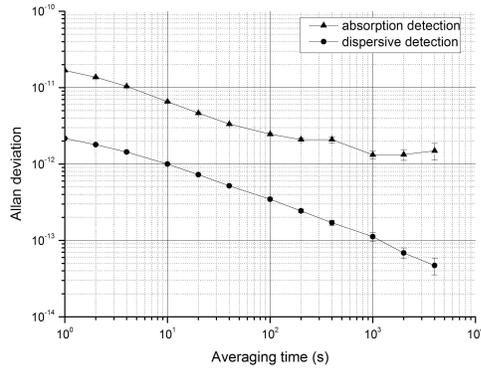}\\
\caption{Allan deviation of the dispersive detection(circle) and the absorption detection(triangle) in POP-type atomic clock. The local oscillator are locked on the Ramsey fringes on Fig. \ref{Ramsey}(b) and (c). The reference frequency standard is H-maser (VCH-1003A,VREMYA-CH) whose stability is $2\times 10^{-13}$ at 1s, and $2\times 10^{-15}$ after $10^4$ s.\label{Allan}}
\end{figure}

\textit{Performance of POP atomic clock with the dispersive detection}---
We have demonstrated the feasibility of the dispersive detection of the clock signal and reported an ultra-high resonance contrast about 80\% above. Now we turn to the POP atomic clock with our novel detection method, in which the microwave interrogation and the detection of the probe light are separated in timing sequence to eliminate light shift. In a conventional POP clock, experimental sequence includes three stages of optical pumping, pulsed operation with microwave fields, and the measurement of the atom number at the upper state of the clock transition, as shown in Fig. \ref{Ramsey}(a). In the modified scheme, however, we replace the conventional absorption detection with the dispersive detection of the transition between $\ket{F=2}$ and $\ket{F'=2}$ to extract the information of the clock signal. Firstly, the atoms in the state $\ket{F=2}$ are transferred to the state $\ket{F=1}$ by optical pumping. Then two microwave pulses with the pulse area $\pi/8$ and an interval of $T=4$ms interrogates the atoms. The atomic population in $\ket{F=2, m_F=0}$ are subsequently detected by a linear-polarized probe light and finally the signal with Ramsey pattern is obtained. Typical clock signal is depicted in Fig. \ref{Ramsey}(b), which exhibits the contrast of about 90\% and the signal-to-noise of 840 (Bandwidth @ 1 kHz). For comparison, we also give the result of the absorption detection of transmitted light without the crossed polarizer in Fig. 3(c). The contrast and the signal-to-noise are 20\% and 130 (Bandwidth @ 1 kHz), respectively. Apparently, the dispersive detection enhances the contrast and the SNR of the clock signal remarkably, which evidently prove the advantage of our detection scheme. The linewidth of the Ramsey pattern at the centre zone are both about 125$\pm$4Hz for Fig. \ref{Ramsey}(b) and (c), and they agree well with the well-known expression $\Delta \nu = 1/2T$. The Allan deviation of the absorption detection and the dispersive detection are given in Fig. \ref{Allan}. At the same stability of temperature, magnetic field, and light intensity, the POP Allan deviation of dispersive detection is improved about an order in our experimental environment. The current stability of the solenoid is about $2\times10^{-6}$ at $1<\tau<10^4$s. The temperature stability is about  $2\times10^{-4}$K at $1<\tau<10^4$s. There is much room for improving the Allan deviation because an operational atomic clock not only depends on the contrast of Ramsey signal, but also depends on other components like temperature stabilization, magnetic field stabilization, and electronics.

\textit{Density effect}---
Hereinbefore we have discussed the dispersive detection when the sample was driven by a weak microwave field. It has been pointed out that density effect plays a prominent role in the dispersive detection when more atoms in the lower state of the probe transition are populated \cite{Pustenlny2005}. In this context, the density effect on atomic clocks should be investigated in detail for a strong driving field on the clock transition. For a weak microwave field, the contribution from the absorption part can be neglected and the dispersive property predominates, however, the absorption cannot be ignored and even becomes dominant when a strong driving field is applied. We have observed the remarkable splitting of the transmission peak when the power of the microwave increases, which are shown in Fig.\ref{split} (b-d). The experimental results agree qualitatively with our theoretical simulation (see Fig.\ref{split}(f-h)).

The splitting effect also appears in the Ramsey fringe of the POP clock. When we increase the two microwave pulses area to $\pi/2$ and maintain their interval of $T=4$ ms, we found that the density effect could lead to the splitting of the transmission peak, and consequently it is expected that the interesting effect produces a narrower linewidth of Ramsey fringes in POP clocks. Experimentally, we only changed the power of the microwave fields to increase the pulse area and remain other parameters unchanged. When two microwave pulses with the same pulse area $\pi/2$ were applied, we observed that the linewidth at the center zone of the signal was about 64$\pm$5 Hz (see Fig. \ref{Ramsey}(d)). It shows obviously that the linewidth is significantly dependent on the pulse area of the microwave fields and thus the atomic population on the upper state of the clock transition. As the microwave pulse area increases, the linewidth become narrower with the contrast almost unchanged (from 87.5\% to 90\%). The profile of Ramsey fringe in Fig. \ref{Ramsey}(d) exhibits a slight asymmetry. In our case, this imperfection originates from residual atomic population after optically pumping, the misalignment between magnetic fields and laser field.

As the dispersive detection is sensitive to the magnetic field, the influence of the magnetic field should be considered carefully. From Eq. (\ref{eq:rho}), one can obtain immediately
\begin{equation}\label{eq:rho-Re-Im}
    \text{Re}(\rho_{\pm}) \propto \frac{\delta_l}{\Gamma^{*2}+\delta_l^2}, \text{Im}(\rho_{\pm}) \propto \frac{1}{\Gamma^{*2}+\delta_l^2},
\end{equation}
where $\delta_l=g_F \mu_B B$ is the level shift produced by the magnetic field, $g_F$ is the Land\`{e} factor, and $\mu_B$ is the Bohr magneton.  In our experiment $\delta_l$ is about 15kHz, and $4\delta_l^2/\Gamma^{*2} \approx 2\times10^{-10} \ll 1$. As the stability of the magnetic field is better than $1\times10^{-3}$, which can be satisfied in our experiment, the stability of magnetic field is not the limitation for the dispersive detection.  Another factor that might affect the performance of atomic clock is the off-resonance excitation of other Zeeman levels. As the level shift is 15kHz, the probability  of the off-resonance excitation is about $P_{\text{15kHz}}=1.8\times10^{-6}$, which is negligible in the detection because the SNR of the experiment is 840.

\textit{Conclusion}---
Our work has demonstrated an intriguing application of Faraday effect in atomic clocks. Ultra-high resonance contrast with high SNR has been realized in double resonance configuration by probing the magneto-optical rotation of polarized plane of incident light. Combined with POP technique, the dispersive detection has remarkable benefit on the clock performance.  In addition, it was confirmed that the density effect can suppress the linewidth of Ramsey fringes. This detection scheme is applicable to other atomic clocks that optically detection of the clock signal is used. With the temperature stability and magnetic stability been improved one or two orders and the phase noise of the microwave generator been suppressed, the stability of our atomic clock would be better than $1\times10^{-13} \tau^{-1/2}$ and these type of atomic clock will become a good substitute of the passive hydrogen maser. We expect that the dispersive detection will have a great effect on future metrology, in various areas ranging from fundamental physics to communication network.

\begin{acknowledgments}
We thank Yongqing Li, Jun Qian, Lin Li and Feng Chen for useful discussions and help. This work was supported by the Major State Basic Research Development Program of China under Grant No.2005CB724507, No.2006CB921202.
\end{acknowledgments}


\newpage

\end{document}